\documentclass[prd, onecolumn, nofootinbib, floatfix]{revtex4}
\usepackage[mathscr]{euscript}
\usepackage{amsmath}
\usepackage{graphicx}
\usepackage{dcolumn}
\usepackage{bm}
\usepackage{epsfig}
\usepackage{amssymb,latexsym,mathrsfs}
\usepackage{graphicx}
\usepackage{color}
\usepackage{hyperref}
\usepackage{float}
\usepackage{diagbox}

\hypersetup{
    colorlinks=true,
    linkcolor=red,
    citecolor=blue,
}
\definecolor{amaranth}{rgb}{0.9, 0.17, 0.31}
\definecolor{purple(munsell)}{rgb}{0.62, 0.0, 0.77}
\definecolor{americanrose}{rgb}{1.0, 0.01, 0.24}
\definecolor{palatinateblue}{rgb}{0.15, 0.23, 0.89}
\definecolor{royalblue(web)}{rgb}{0.25, 0.41, 0.88}
\definecolor{hanpurple}{rgb}{0.32, 0.09, 0.98}
\definecolor{beaublue}{rgb}{0.74, 0.83, 0.9}
\definecolor{carminered}{rgb}{1.0, 0.0, 0.22}
\definecolor{brightpink}{rgb}{1.0, 0.0, 0.5}
\definecolor{vividviolet}{rgb}{0.62, 0.0, 1.0}
\hypersetup{ linktoc=all,
    colorlinks, linkcolor={palatinateblue},
    citecolor={brightpink}, urlcolor={amaranth}}

\newcommand{\changeurlcolor}[1]{\hypersetup{urlcolor=#1}}
\newcommand{\be}{\begin{equation}}
\newcommand{\ee}{\end{equation}}
\newcommand{\bs}{\begin{split}}
\newcommand{\bea}{\begin{eqnarray}}
\newcommand{\eea}{\end{eqnarray}}

\newcommand{\la}{\langle}
\newcommand{\ra}{\rangle}
\newcommand{\w}{\omega}
\newcommand{\ep}{\epsilon}

\newcommand{\bes}{\begin{subequations}}
\newcommand{\ees}{\end{subequations}}

\begin{document}

\title{Late time approach to Hawking radiation: terms beyond leading order}
\author{Paul~R.~Anderson${}^{1}$}
\email{anderson@wfu.edu}
\author{Raymond D. Clark${}^{1}$}
\author{Alessandro~Fabbri${}^{2}$}
\email{afabbri@ific.uv.es}
\author{Michael R.R. Good${}^{3}$}
	\email{michael.good@nu.edu.kz}
\affiliation{${}^1$Department of Physics, Wake Forest University, Winston-Salem, North Carolina 27109, USA}
\affiliation{${}^2$Departamento de F\'isica Te\'orica and IFIC, Centro Mixto Universidad de Valencia-CSIC, C. Dr. Moliner 50, 46100 Burjassot, Spain}
\affiliation{${}^3$Department of Physics and ECL, Nazarbayev University, Astana, Kazakhstan}

\begin{abstract}

Black hole evaporation is studied using wave packets for the modes.  These allow for approximate frequency and time resolution.  The leading order late time behavior gives the well known Hawking radiation that is independent of how the black hole formed. The focus here is on the higher order terms and the rate at which they damp at late times.  Some of these terms carry information about how the black hole formed.  A general argument is given which shows that the damping is significantly slower (power law) than what might be naively expected from a stationary phase approximation (exponential).  This result is verified by numerical calculations in the cases of 2D and 4D black holes that form from the collapse of a null shell.

\end{abstract}

\date{\today}

\maketitle



Hawking's seminal work showed that particle production occurs when matter collapses to form a black hole and that at late times the particles are produced
in a thermal spectrum (modulo the gray body factor) at a temperature that is related to the surface gravity at the event horizon of the black hole~\cite{Hawking:1974sw}.
This gives rise to the well known information issue~\cite{Hawking-PRD}  which involves the apparent loss of information during the process of formation and evaporation of the black hole.

Although the leading order late time thermal radiation has been thoroughly studied in many different ways, much less attention has been paid to the early time nonthermal radiation which at late enough times becomes negligible compared with the thermal radiation.  In~\cite{Good:2016MRB} a study was done on the time dependence of the particle
production that occurs due to a massless minimally coupled scalar field when a spherically symmetric null shell collapses to form a black hole in both two and four spacetime dimensions.
An exact correspondence \cite{MG14one,MG14two,Good:2016LECOSPA,Good:2016MRB} between the particle production that occurs due to the collapse of the shell and
that which occurs in 2D for a mirror undergoing a particular trajectory in flat space was shown.

In this paper we study the late time behavior of the particle production process, focusing on the late time behavior of the terms that are beyond leading order.  After deriving a general form for the equations that describe the particle production process, we show for a specific example in 2D that a stationary phase approximation gives the late time leading order results found by Hawking with correction terms that are exponentially damped in time.  We then show in general that the true next to leading order (and next to next to leading order, etc.) terms are power law damped at late times.

For simplicity, the derivation we give uses the form of the Bogolubov coefficients for a massless minimally coupled scalar field for a 2D asympotically flat spacetime in which a Schwarzschild black hole forms from collapse.  However, the derivation can be generalized immediately and trivially to the case of 4D spherically symmetric collapse and almost certainly to the general case of 4D collapse to form a black hole in an asymptotically flat spacetime.

We begin by noting that in the case of collapse to form a black hole one can expand the quantum field in terms of modes that correspond to the natural {\it in} vacuum state.  For an asymptotically flat spacetime these are positive frequency at past null infinity, $\mathscr{I}^-$, and regular at the origin, $r = 0$, inside the collapsing matter.  (This is a necessity in 4D.)  We denote these modes by $f^{\rm in}_\w$.  For asymptotically flat spacetimes the modes in the {\it out} vacuum state that we are concerned with are positive frequency at future null infinity, $\mathscr{I}^+$,  and vanish on the future horizon $H^+$.  We denote them by $f^{\rm out}_\w$.  Note that they do not form a complete set.   Then in the usual way we write one of these modes in terms of the complete set of {\it in} modes as follows:
\be f^{\rm out}_\w = \int_0^\infty d \w' \left[ \alpha_{\w \w'} \, f^{\rm in}_{\w'} + \beta_{\w \w'} \, f^{\rm in \; \*}_{\w'} \right]. \ee
The Bogolubov coefficients are computed using the usual scalar product~\cite{Birrell:1982ix} with the result $\alpha_{\w\w'} = (f^{\rm out}_{\w}, f^{\rm in}_{\w'})$, $\beta_{\w  \w'} = -(f^{\rm out}_{\w }, f^{\rm in \; *}_{\w'}) $.
The average number of particles produced at frequency $\w$ is,
\be \la {\rm in}| N^{\rm out}_{\w} |{\rm in} \ra = \int_0^\infty d \w'  \, |\beta_{\w \w'}|^2 \;. \label{Nw} \ee
This number is divergent if backreaction effects are not taken into account since the black hole then radiates forever.

To investigate the time-dependence of the particle production process Hawking used wave packets for the {\it out} modes,
\be f^{\rm out}_{j n} = \frac{1}{\sqrt{\epsilon}}\int_{j \epsilon}^{(j+1) \epsilon} d \omega \; e^{i 2 \pi n \ep^{-1} \omega}\; f^{\rm out}_{\omega}  \;, \ee
with $n$ an integer and $j$ a nonnegative integer.  Note that near future null infinity, $\mathscr{I}^+$, $f^{\rm out}_\w \sim  e^{- i \w (t_s - r_*)}$, with $t_s$ the usual time coordinate in Schwarzschild spacetime and $r_* \equiv r + 2 M \log[(2M)^{-1} (r-2M)] $.  Thus the contribution to the integral is largest if $u_s = t_s-r_* = 2 \pi n/\ep$.  A packet with index $j$ covers the range of frequencies
$j \ep \le \w \le (j+1)\ep$, and one with index $n$ covers the approximate time range
$(2 \pi n - \pi)/\ep \lesssim u_s \lesssim (2 \pi n + \pi)/\ep$.

One can also make packets out of the Bogolubov coefficients \be \beta_{j n \omega^{'}} = \frac{1}{\sqrt{\epsilon}}\int_{j \epsilon}^{(j + 1) \epsilon} d \omega \; e^{i 2 \pi n \epsilon^{-1} \w}  \beta_{\omega \omega^{'}}.  \ee
It is useful to write the Bogolubov coefficients in the following form
\be \beta_{\w \w'} = (\w')^{-1/2} e^{-i (\w + \w') v_H} e^{-i \kappa^{-1} \w \log \left( \kappa^{-1} \w' \right)} B(\w, \w') \;.  \label{B-def} \ee
Here, as discussed in~\cite{Hawking:1974sw}, $v_H$ is the latest time that a left moving radial null geodesic can leave at $\mathscr{I}^{-}$,  pass through the center of the collapsing object, and reach $\mathscr{I}^{+}$, while $\kappa$ is the surface gravity of the black hole.  Then the number of particles in a packet is
\bes \bea      N_{j n}  &\equiv& \frac{1}{\ep} \int_{0}^{\infty}d \omega^{'} \mid \beta_{j n \omega^{'}} \mid ^2  =    \frac{1}{\ep} \int_{0}^{\infty}\frac{d\w^{'}}{\w'} \int_{j \ep}^{(j+1) \ep} d \w_1 \int_{j \ep}^{(j+1) \ep}  d \w_2
 e^{i A (\w_1-\w_2)} B(\w_1, \w') B^{*}(\w_2, \w')\ ,  \label{triple}  \\
 A &\equiv&  2 \pi n\epsilon^{-1}- v_H -\kappa^{-1} \log \left( \kappa^{-1} \w' \right)   \;.  \label{A-def} \eea \ees

A stationary phase approximation can be used to determine the leading order behavior with the stationary phase point,
\be   \w'_{stph}  = \kappa \, e^{2 \pi n \kappa \ep^{-1} - \kappa v_H} \;. \label{wstph} \ee
At late times $ \w'_{stph} \gg \kappa $,
\be B(\w,\w') =  B_H(\w) + B_1(\w,\w')  \;,  \label{B-expansion} \ee
with $B_H$ the asymptotic value found by Hawking and $B_1 \to 0$ in the limit $\w' \to \infty$.\footnote{The form~\eqref{B-expansion} can be used to  illustrate the thermal behavior at late times in other cases as well such as the 2D mirror trajectories studied in~\cite{Davies:1976hi,Davies:1977yv,Carlitz:1986nh, Hotta:1994ha, Good:2012cp,Good:2013lca,Good:2015nja, MG14one,MG14two,Good:2016LECOSPA,Good:2016MRB, horizonless, paper1, paper2, paper3, Hotta:2015yla, harvest,Good:2016HUANG}. For those trajectories one finds that $B_H$ is also given by~\eqref{BH2D}.  Substituting into~\eqref{B-def} gives  $|\beta^{H2D}_{\w,\w'}|^2 = \frac{1}{2 \pi \kappa \w'} \; \frac{1}{e^{2 \pi \w/\kappa}-1} $.}
Using~\eqref{B-expansion} in~\eqref{triple} and keeping only the first term gives Hawking's result which is independent of $n$.
For the case of a collapsing null shell in 2D one finds~\cite{Good:2016MRB,Massar:1995im}
\bes \bea B^{2D}(\w,\w') &=& \frac{B^{2D}_H(\w)}{(1+ \w/\w')^{1+ i \kappa^{-1} \w}} \;, \label{B2D} \\
 B^{2D}_H(\w) &=& -\frac{\sqrt{\w}}{2 \pi \kappa} e^{- \frac{\pi \w}{2 \kappa}} \Gamma\left( i \kappa^{-1} \w \right) \;. \label{BH2D} \eea \ees
Substitution into~\eqref{triple} and evaluation at the stationary phase point~\eqref{wstph} shows that in this case the next to leading order terms are exponentially damped in $n$.
This analysis might lead one to believe that the distribution of particles approaches the thermal one 
predicted by Hawking  exponentially in time.
However, as shown next, for the general case of collapse to form a black hole in an asymptotically flat spacetime, there are contributions to $N_{jn}$ that are damped like inverse powers of $n$ at late times.

To see this, first assume $2 \pi n \kappa \ep^{-1} \gg 1$ and then break the integral over $\w'$ into two parts with the first part going from $\w' = 0$ to $ \Lambda$ and the second part going from $\w' = \Lambda$ to $ \infty$.  The cutoff $\Lambda$ should be chosen so it is in the range $(j+1) \epsilon \ll \Lambda \ll \w'_{stph} $. This will always work for fixed $j$ and $\epsilon$ if large enough values of $n$ are used.
Then for the second term the same triple integral with $B(\w,\w') \to B_{H}(\w)$ is subtracted off and added back on with the result
\bes \begin{eqnarray}   N_{j n}   &=&  N_1 + N_2 + N_3 \nonumber \\
   N_1 &=&   \frac{1}{\ep}\int_{0}^{\Lambda} \frac{d \w^{'}}{\w^{'}} \int_{j \ep}^{(j+1) \ep} d \w_1 \int_{j \ep}^{(j+1) \ep} d \w_2   e^{i 2 \pi n \epsilon^{-1}(\w_1-\w_2)} \beta_{\w_1 \w'} \beta^{*}_{\w_2 \w'}\ ,  \\ & &  \nonumber \label{N1} \\
  N_2 &=& \frac{1}{\ep}\int_{\Lambda}^{\infty}\frac{d \w^{'}}{\w^{'}} \int_{j \ep}^{(j+1) \ep} d \w_1 \int_{j \ep}^{(j+1) \ep} d \w_2 e^{i A (\w_1-\w_2)} \, \left[ B(\w_1,\w') B^{*}(\w_2,\w')
            - B_H(\w_1) B_H^{*}(\w_2) \right]\ ,    \label{N2}\\
    N_3 &=& \frac{1}{\ep}\int_{\Lambda}^{\infty}\frac{d \w^{'}}{\w^{'}} \int_{j \ep}^{(j+1) \ep} d \w_1 \int_{j \ep}^{(j+1) \ep} d \w_2               e^{i A(\w_1-\w_2)} B_H(\w_1) B_H^{*}(\w_2)\ . \label{N3}
   \eea \label{N1N2N3} \ees
Note that $N_{j n}$ is independent of the value of $\Lambda$.

The leading and next to leading order behaviors for large values of $n$ come from $N_3$.  To see this break the integral over $\w'$ in $N_3$ into two parts such that $\int_\Lambda^\infty d\w' = \int_0^\infty d\w' - \int_0^\Lambda d\w'$.
Integrating the first term over $\w'$ and then $\w_2$, and dropping the subscript on $\w_1$  gives the leading order late time behavior found by Hawking~\cite{Hawking:1974sw}
\be N_{3a} = N^H_{jn} = \frac{2 \pi \kappa}{\ep} \int_{j \ep}^{(j+1)\ep} d \w |B_H(\w)|^2 \;, \label{N3a2}   \ee
which is independent of $n$.

The next to leading order term is obtained from the second term for $N_3$ by integrating the integrals over both $\w_1$ and $\w_2$ by parts and noting that since
$j$ and $n$ are integers, $e^{i 2 \pi n (j+1)} = e^{i 2 \pi n j} = 1$.
The result is
\be N_{3b} = -\frac{\kappa}{2 \pi n} \left[ |B_H((j+1)\ep)|^2 + |B_H(j \ep)|^2 \right] + O(n^{-2}) \;.  \label{N3b} \ee
It is easy to see by integrating the integrals over $\w_1$ and $\w_2$ in $N_1$ and $N_2$ by parts that for large enough values of $n$, $N_{1,2} = O(n^{-2})$.

It is important to note that some of the terms in $N_1$ and $N_3$ and all of the terms in $N_2$ that are of order $n^{-2}$ at large $n$ depend on the cutoff $\Lambda$.  However, it is clear from~\eqref{N1N2N3} that the sum of these terms cannot depend on $\Lambda$.  Therefore it is the terms that do not depend on $\Lambda$ that we focus on.

We have checked these results for the case of particle production due to a massless minimally coupled scalar field that occurs when a black hole forms by the collapse of a null shell in 2D and 4D.  In the latter case the shell is spherically symmetric.  Previous studies of particle production when a null shell collapses in 4D to form a black hole were done in~\cite{Vilkovisky,Massar:1995im,Fabbri:2005mw,Good:2016MRB}~\footnote{In these studies the effective
potential in the mode equation for the scalar field was ignored (with the exception in some cases of the gray body factor) allowing for analytic computation of the Bogolubov coefficients.}.

In the 2D case as previously mentioned the Bogolubov coefficients were computed exactly in~\cite{Good:2016MRB,Massar:1995im} with the result~\eqref{B2D}.
Substitution of~\eqref{BH2D} into~\eqref{N3a2}  gives for $\frac{j \ep}{\kappa} \gtrsim 1$ and small values of $\frac{\ep}{\kappa}$

\be  N_{3a} = N^H_{j n}=  \frac{1}{e^{2 \pi \w_j/\kappa}-1}   + O(\frac{\ep^2}{\kappa^2}) \;, \label{N3a2D}    \ee
 where
  $\w_j = (j+\frac{1}{2}) \ep $.
  Substitution of~\eqref{BH2D} into~\eqref{N3b} gives
\be N_{3b} = \frac{1}{4 \pi^2 n} \left[ \frac{1}{e^{2 \pi (j+1) \ep \kappa^{-1}}-1} + \frac{1}{e^{2 \pi j \ep \kappa^{-1}}-1} \right] + O(n^{-2}) \;. \label{Njn-2D-subleading} \ee
 The quantities $n^2 N_1$, $n^2 N_2$, and $n N_{3b}$ are plotted in Fig.~(\ref{fig-2D-next}). As predicted they each approach a constant value in the large $n$ limit.  For $n N_{3b}$ the constant can be obtained by multiplying~\eqref{Njn-2D-subleading} by $n$ and taking the limit $n \to \infty$.
\begin{center}
\begin{figure}[ht]
\includegraphics [totalheight=0.25\textheight]
{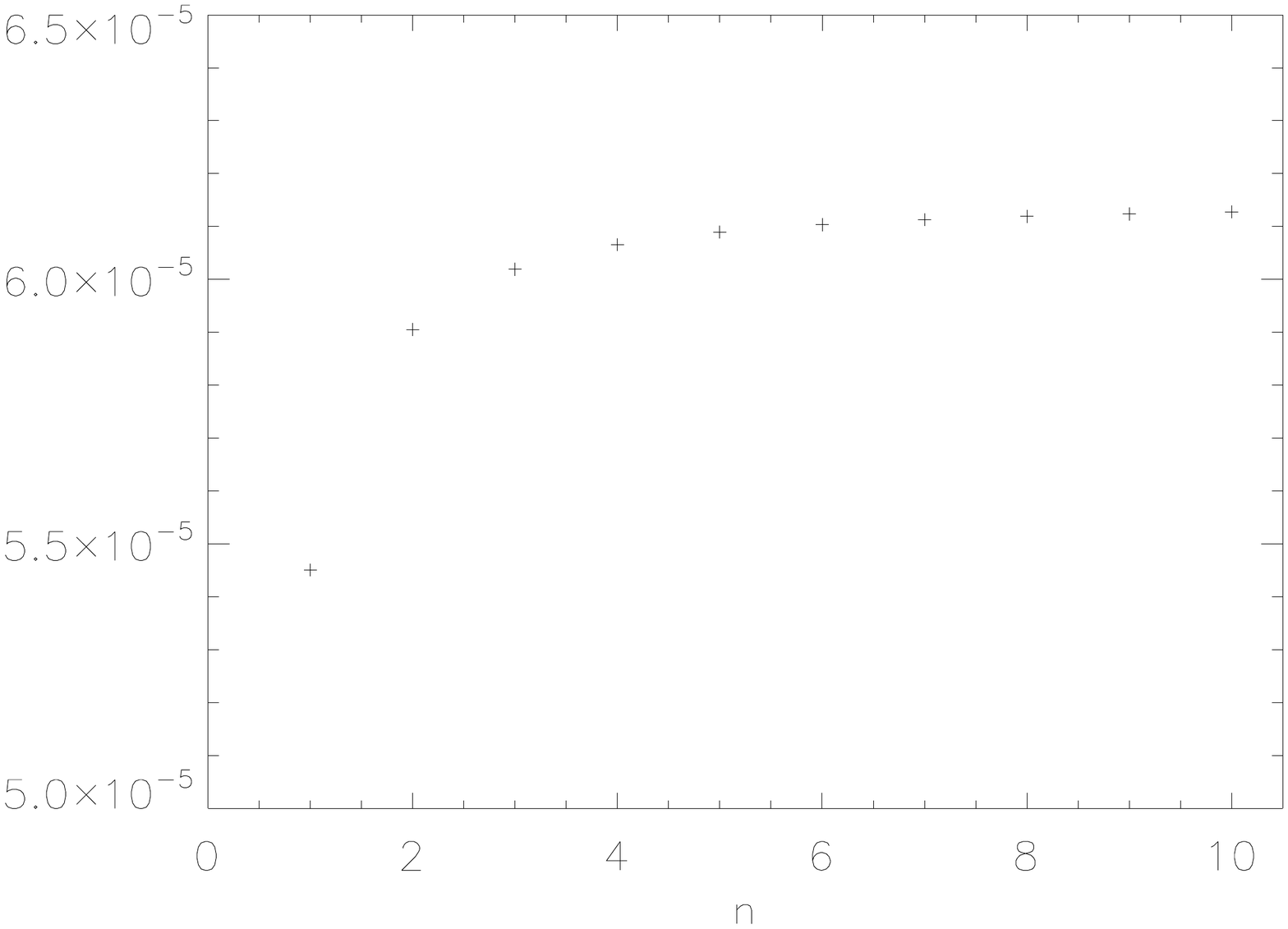}
\includegraphics [totalheight=0.25\textheight]
{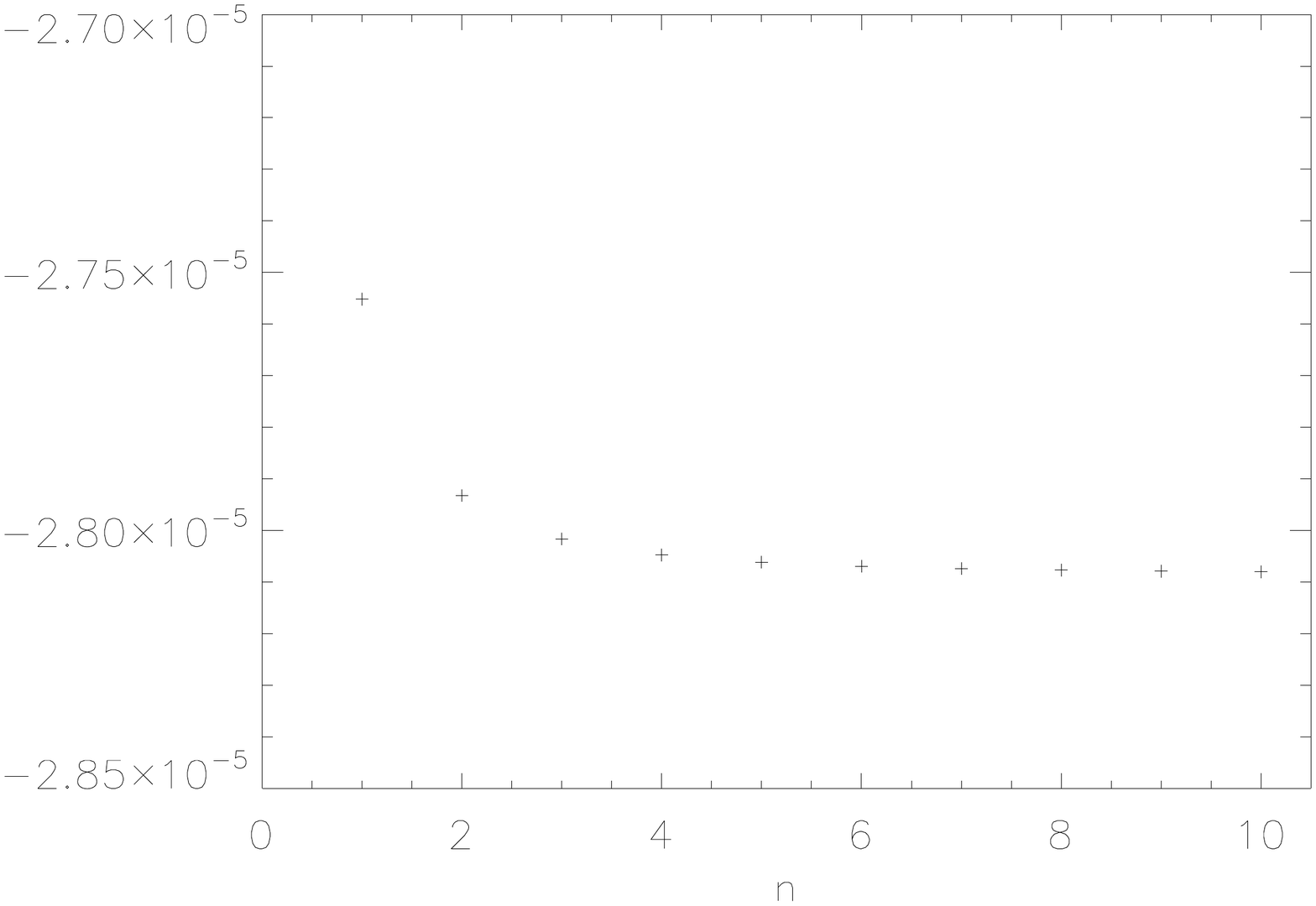}
\includegraphics [totalheight=0.25\textheight]
{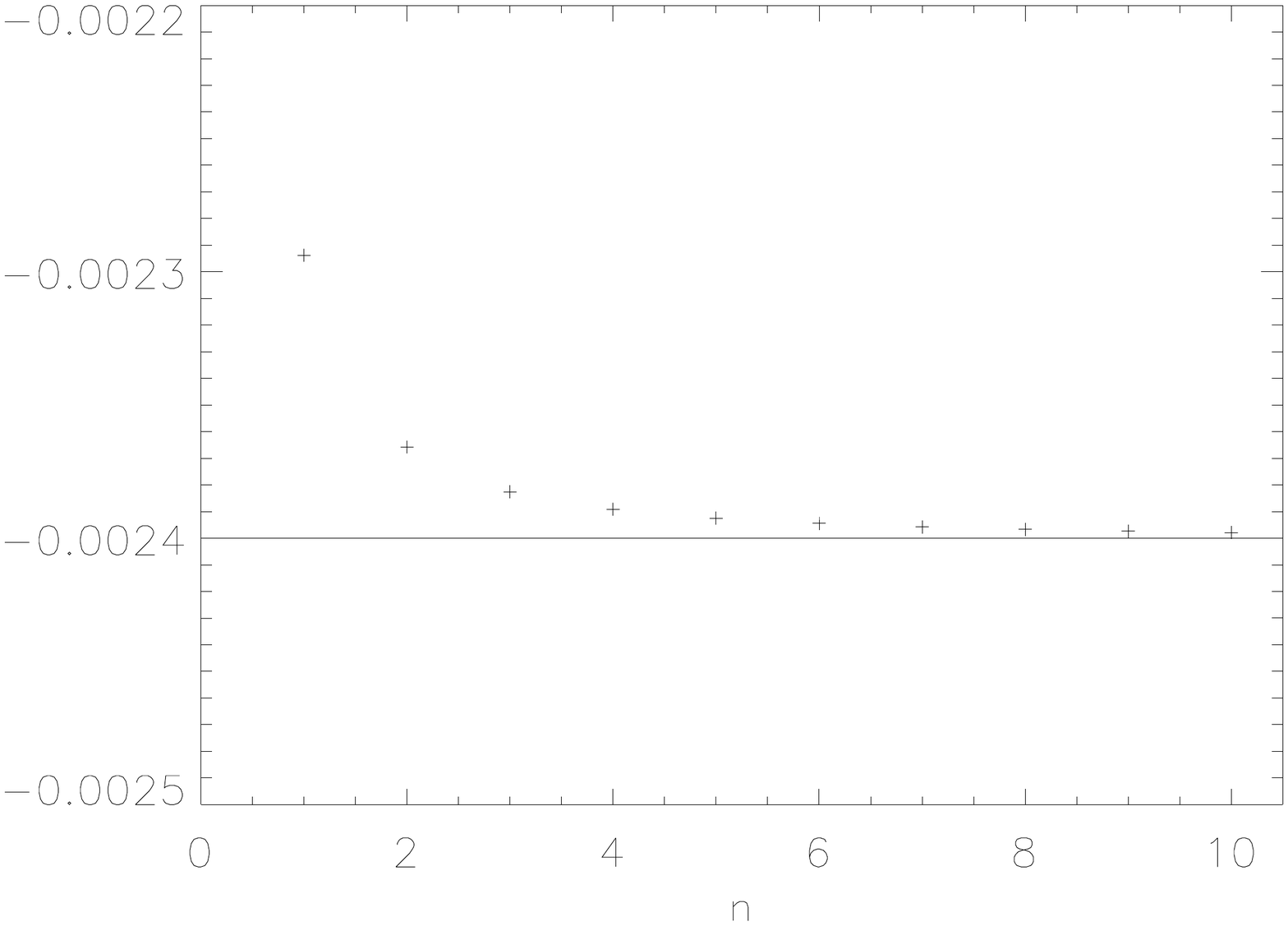}
\caption{The quantity $n^2 N_1 $ is plotted versus $n$ in the upper left hand plot and the quantity $n^2 N_2$ is plotted in the upper right hand plot for the case of the 2D collapsing null shell with $M = 1$, $j = 1$, $\ep = 0.1$, $v_0 = 0$, and $\Lambda = 10$.  The quantity $n N_{3b}$ is plotted in the lower plot.  The solid line is the limit $n \to \infty$ of $n N_{3b}$ which can be obtained from~\eqref{N3b} and which is equal to $-0.00239997 $.
 }
\label{fig-2D-next}
\end{figure}
\end{center}

In 4D 
the relevant Bogolubov transformation is
\be
f^{\rm out}_{\w \ell m} = \sum_{\ell' m'} \int_0^\infty d \w'
\left[ \alpha_{\w \ell m \w' \ell' m'} f^{\rm in}_{\w' \ell' m'}
+ \beta_{\w \ell m \w' \ell' m'} f^{{\rm in} \, *}_{\w' \ell' m'} \right] \;,
\label{Bog-4D}
\ee
where $l,m$ are the usual angular momentum parameters.
Using the Cauchy surface consisting of $\mathscr{I}^-$ for $v_0 \le v < \infty$ plus the trajectory of the null shell, $v = v_0$~\cite{Good:2016MRB}, 
we find 
that for the large $\w'$ expansion~\eqref{B-expansion} 
\be
B_H(\w)  = \frac{B^{2D}_H(\w)}{F_L}  \;, \label{BH4D}
\ee
 where $\frac{1}{|F_L|^2}$ is the gray-body factor \cite{rigorous-results}, and the next to leading order term $B_1$ is more involved than in the 2D case and depends on two of the scattering coefficients and the angular momentum parameter $l$.
 The details will be presented elsewhere.

 We have numerically computed $N^H_{jn}$ for the case $M = 1$, $j = 1$, $\ep = 0.1$, and $v_0 = 0$.  We have also numerically computed in a direct way the value of $N_3$ for various values on $n$ when $\Lambda = 10$.  In Fig~\ref{fig-4D}  the quantity $n N_{3b}$ is plotted.  As predicted it approaches a constant value in the large $n$ limit. The constant can be obtained by multiplying~\eqref{N3b} by $n$ and taking the limit $n \to \infty$. 
\begin{center}
\begin{figure}[ht]
\includegraphics [totalheight=0.3\textheight]
{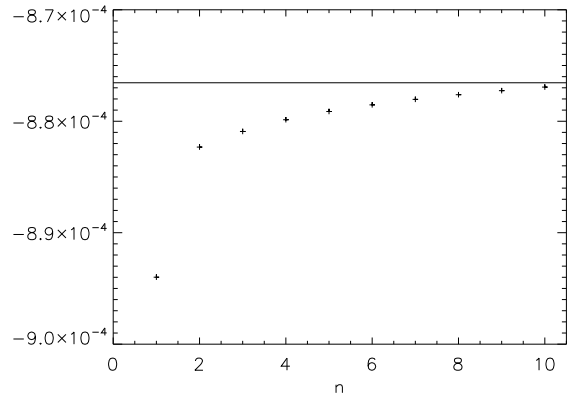}
\caption{The quantity $n N_{3b}$ is plotted for the case of the 4D collapsing null shell with $M = 1$, $j = 1$, $\ep = 0.1$, $v_0 = 0$, and $\Lambda = 10$.  The solid line is the limit $n \to \infty$ of $n N_{3b}$ which can be obtained from~\eqref{N3b} and which is equal to $-0.00087655 $.
}
\label{fig-4D}
\end{figure}
\end{center}
We have shown that the contribution to the particle production from the next to leading order term goes like $n^{-1}$, with $n$ related to the time, when wave packets are used for the {\it out} modes.  This was unexpected since a stationary phase approximation would seem to predict that higher order terms are exponentially damped in $n$.

Interestingly the next to leading order term is still local since it depends on the leading order contribution to the Bogolubov coefficient $\beta_{\w \w'}$ in a large $\w'$ expansion and, as Hawking showed, this is independent of the details about how the black hole formed.  However, at the next order, $n^{-2}$, and for all higher orders $n^{-a}$ for $a >2$, there are cutoff independent contributions to $N_1$ in~\eqref{N1} which are clearly nonlocal since they depend on much smaller values of $\w'$ that require a full numerical computation of $\beta_{\w \w'}$ over the entire trajectory of the null shell
as well as the contribution to $\beta_{\w \w'}$ from past null infinity.  Thus these terms carry information about how the black hole formed.

Our findings that at late times the terms beyond leading order in the Hawking effect have a powerlaw rather than an exponential decay and that the first nonlocal terms emerge at order $n^{-2}$ put the information loss issue in a new perspective since the emitted particles in the black hole evaporation process are sensitive to the details of the collapsing matter for a significantly longer period of time than would have been naively expected from a stationary phase approximation.  Of course for a macroscopic black hole one would still expect that the approach to a thermal state would occur well before a significant amount of evaporation has occurred.

\acknowledgments
P.R.A. thanks Amos Ori, Adam Levi, Ted Jacobson, and Abhay Ashtekar for helpful conversations, comments, and questions, and Adam Levi for sharing some of his numerical data.  We would also like to thank Jose Navarro-Salas for helpful comments on the manuscript.  This work was supported in part by the National
Science Foundation under Grants No. PHY-1505875 and PHY-1912584 to Wake Forest University.
R.D.C. acknowledges financial support from the Wake Forest University URECA Center.
A.F. acknowledges partial financial
support by the Spanish Ministerio de Econom´ıa, Industria y Competitividad Grant
No. FIS2017-84440-C2-1-P, the Generalitat Valenciana Project No. SEJI/2017/042 and the Severo Ochoa Excellence
Center Project No. SEV-2014-0398.
M.R.R.G acknowledges funding from state-targeted program ``Center of Excellence for Fundamental and Applied Physics" (BR05236454) by the Ministry of Education and Science of the Republic of Kazakhstan and support by the ORAU FY2018-SGP-1-STMM Faculty Development Competitive Research Grant No. 090118FD5350 at Nazarbayev University.  Some of the plots were made using  the WFU DEAC cluster; we thank the WFU Provost's Office and Information
Systems Department for their generous support.

\newpage

\end{document}